\documentclass[10pt,conference,compsocconf,letterpaper]{IEEEtran}
\makeatletter

\def\ps@headings{%

\def\@oddhead{\mbox{}\scriptsize\rightmark \hfil \thepage}%

\def\@evenhead{\scriptsize\thepage \hfil \leftmark\mbox{}}%

\def\@oddfoot{}%

\def\@evenfoot{}}

\makeatother

\ifCLASSINFOpdf
\else
\fi
\usepackage{graphicx}
\usepackage{algorithm}
\usepackage{subfigure}
\usepackage{multirow}
 \usepackage{graphicx}
  \usepackage{algorithm}
\usepackage{setspace}
\usepackage{subfigure}
\usepackage{array}
\newcommand{\PreserveBackslash}[1]{\let\temp=\\#1\let\\=\temp}
\newcolumntype{C}[1]{>{\PreserveBackslash\centering}p{#1}}
\newcolumntype{R}[1]{>{\PreserveBackslash\raggedleft}p{#1}}
\newcolumntype{L}[1]{>{\PreserveBackslash\raggedright}p{#1}}
\usepackage{amsfonts}
\usepackage{amsmath}
\usepackage{url}
\begin{document}
%
\title{\LARGE A QoS Guarantee Strategy for Multimedia Conferencing based on Bayesian Networks}
%
%
%

\author{
Junfei Huang and Guochu Shou\\
State Key Laboratory of Networking and Switching Technology, \\
Beijing University of Posts and Telecommunications, Beijing 100876, P. R. China.\\
Email: junfei.huang@outlook.com, gcshou@bupt.edu.cn.
       }

\maketitle

\begin{abstract}
Service Oriented Architecture (SOA) is commonly employed in the design and implementation of web service systems. The key technology to enable media communications in the context of SOA is the Service Oriented Communication. To exploit the advantage of SOA, we design and implement a web-based multimedia conferencing system that provides users with a hybrid orchestration of web and communication services. As the current SOA lacks effective QoS guarantee solutions for multimedia services, the user satisfaction is greatly challenged with QoS violations, e.g., low video PSNR (Peak Signal-to-Noise Ratio) and long playback delay. Motivated by addressing the critical problem, we firstly employ the Business Process Execution Language (BPEL) service engine for the hybrid services orchestration and execution. Secondly, we propose a novel context-aware approach to quantify and leverage the causal relationships between QoS metrics and available contexts based on Bayesian networks (CABIN). This approach includes three phases: (1) information discretization, (2) causal relationship profiling, and (3) optimal context tuning. We implement CABIN in a real-life multimedia conferencing system and compare its performance with existing delay and throughput oriented schemes. Experimental results show that CABIN outperforms the competing approaches in improving the video quality in terms of PSNR. It also provides a one-stop shop controls both the web and communication services.
\end{abstract}

\maketitle


\IEEEpeerreviewmaketitle
\section{Introduction}
Service Oriented Architecture (SOA) is a collection of development principles for distributed service systems [1-2]. Web service is a major technology to implement SOA and it enables loosely coupled SOA and interoperable solutions across heterogeneous platforms and systems [3]. Service Oriented Communications (SOC), which uses the web services to enable communications, has already become a hot research issue in both industry and research domains [4-5]. The idea of using web services to support multimedia services is attractive as it can be easily integrated with other SOA-based business systems [6-7]. Furthermore, the web-based multimedia services enjoys better scalability, manageability and less cost for the development of multimedia systems [8]. Recent advances in web services technology have made it practical to enable web-based multimedia communications. With the convergence of Internet and Telecom networks, the integration of web and communication services is becoming an inevitable trend for the service providers [9-10].
\begin{figure*}
\centering
 \includegraphics[width=0.85\textwidth,keepaspectratio]{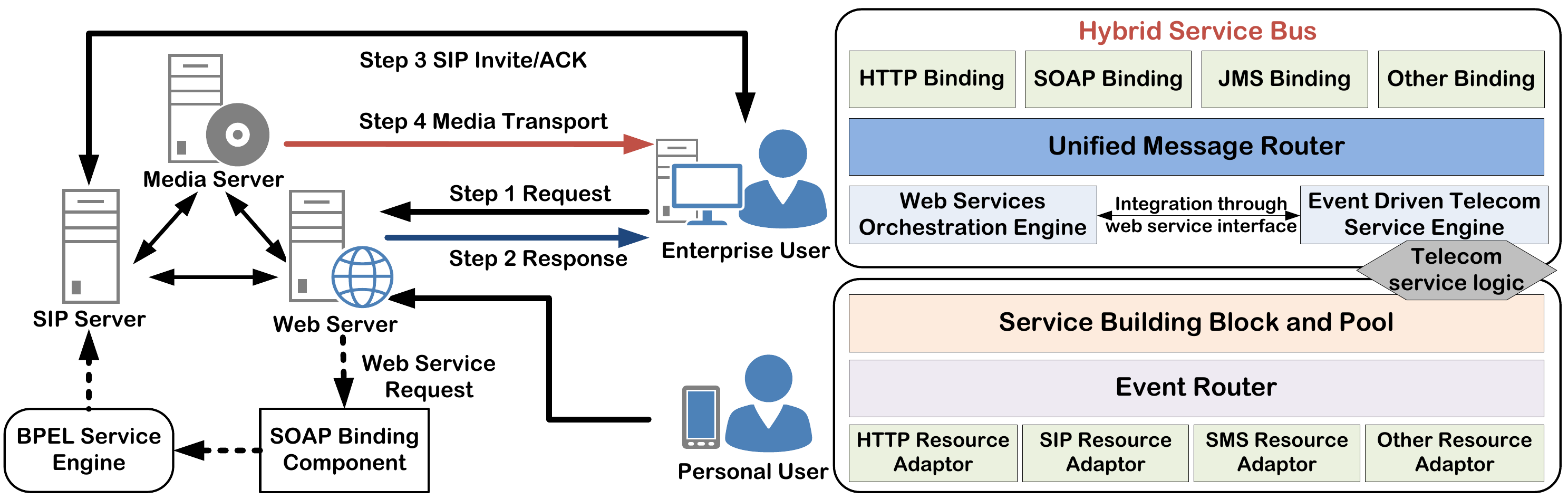}
 \caption{Illustration of the hybrid services orchestration and execution environment of the multimedia conferencing system. The hybrid service bus is based on the Java Business Integration (JBI) standard and acts as the hub for integrating web and communication services through messaging, event handling, business process management, etc. The binding components are used to send and receive messages via different transport protocols. }\label{9}
\end{figure*}

Multimedia conferencing serves as the basis of many prevalent applications, e.g., audio/video conferencing, multi-player online games and remote learning [11]. The traditional multimedia conferencing systems are expensive and challenging to develop and maintain since they are typically monolithic systems [8]. Motivated by the principles of SOA, we have designed and implemented a web-based multimedia conferencing system [12] (see Fig. 1) based on reusable components, e.g., the services orchestration engine and telecom service engine. However, the system users' satisfaction is still greatly challenged with QoS violations (e.g., low video PSNR and long playback delay) because:
\begin{enumerate}
  \item On one hand, the multimedia communication services are characterized by stringent performance requirements, e.g., high-throughput, low-latency and error-resilience; On the other hand, the current SOA stack lacks of effective QoS guarantee mechanisms for multimedia services [3,6,8].
  \item It is difficult to trace and identify the causal relationships between QoS metrics and available contexts\footnote{By the term `contexts', we mean the available system information of application-level metrics (e.g., the user-perceived playback delay and stored video in the playback buffer) or network-level parameters (e.g., available bandwidth, packet loss rate or round trip time) obtained through specific monitoring or sensing mechanisms.}. Furthermore, even if the relationship is determined, it is still a challenging task to carry out context adaption strategies due to the heterogeneity of conferencing participants' networks environments, hardware, QoS requirements, etc.
  \item The number of tunable contexts is very limited at the Internet Service Provider (ISP) side, as many parameters are under the control of Telecom operators and hardware manufacturers [13].
\end{enumerate}

To address the first problem, the BPEL (Business Process Execution Language) service engine, which can describe and organize the services in a standardized manner, is employed for the hybrid services orchestration. Moreover, as the current data transmission protocols (e.g., SOAP and Restful) for web services are not capable of supporting the multimedia streaming well, the RTP (Real-time Transmission Protocol) is employed for media delivery.

On tackling the remaining two problems, the end-to-end QoS management [14] has proved to be a promising solution and the existing schemes can be generally divided into two branches: active probing (e.g., Keynote [15] and Gomez [16]) and passive monitoring (e.g., the CEM [17]). A major problem with both the active and passive schemes is that they either focus on the network level or application level metrics and adapt the tunable contexts accordingly. Therefore, many problems arise from this isolated working manner, e.g., the bandwidth-aware strategies aim at maximizing the throughput of the applications while the upper layer application can not directly benefit form that as it is a typical Constant Bit Rate (CBR) one. In this paper, we propose to analyze and exploit the relationships between the target QoS metrics and available contexts in a comprehensive way, i.e., examining the intrinsic causality between QoS metrics and contexts so as to find out the optimal adaption strategies. This research makes the following contributions:
\begin{enumerate}
  \item We design and develop a SOA-based multimedia conferencing system that employs a novel hybrid web and communication services orchestration mechanism. Based on the implemented conferencing system, we conduct measurements of the QoS metrics and contexts.
  \item We profile the causal relationships between the QoS metrics and contexts through the measurements. A context-aware approach based on Bayesian networks (CABIN) is proposed to quantify and leverage such relationships and it includes three key phases: (1) information feedback and training, (2) QoS-to-context mapping, and (3) optimal context adaption.
  \item We evaluate the performance of CABIN in the multimedia conferencing system. Experimental results show that: (1) CABIN improves the average video PSNR (Peak-Signal-to-Noise Ratio) by up to $3.13$ and $4.16$ dB compared to the throughput oriented (TOR) [25-26] and delay oriented (DON) [27-28] schemes, respectively; (2) CABIN exhibits good performance in preventing the playback buffer starvation and reduces the average playback delay by up to $21.3$ms compared to the TON scheme; (3) CABIN effectively exploits the available capacity for maximizing the video streaming throughput and increases the mean value by up to $77.5$Kbps compared to the DON sheme.
\end{enumerate}

The remainder of this paper is structured as follows. In Section 2, we briefly review the related work and discuss our research motivation. In Section 3, we describe the overall design and key components of CABIN in detail. The implementation and evaluation of the proposed CABIN in real multimedia conferencing system are provided in Section 4. Conclusions are given in Section 5.
\section{Related Work and Research Motivation}
In this section, we firstly review the related work to this paper. Then, we discuss our research motivation from the practically observed contexts and QoS metrics.
\subsection{Related Work}
The related work to this paper can be classified into two categories: (1) service oriented communications, and (2) end-to-end QoS management solutions.

Decneut et al. [7] discuss the use of multimedia Web services in heterogeneous multimedia environments. The work is targeted at the distribution of multimedia data for a variety of client configurations and network conditions. The transmission of multimedia content is achieved using the now obsolete DIME attachment scheme over HTTP, TCP or UDP channels. A multimedia Web service is only allowed to send multimedia content within one single package. That means streaming data is not supported, or it has to be transmitted as if the packets are a big chunk of multimedia content. Zhang et al. [33] propose a framework based on SOAP-oriented component for multimedia Web service. In this work a multimedia Web service is separated into a control flow and a data flow. The control flow is transmitted via ordinary Web services whereas the data flow is managed by the existing multimedia infrastructure. The study proposed a SOAP enhancement so that multimedia data can be alternatively transmitted using SOAP. The enhancement includes the use of `boxcarring' and batching of messages. However, since the transfer of each part of a big chunk of multimedia data requires a separate Web service request, the transfer may be too costly and inefficient when streaming multimedia is transmitted using one separate request per packet.

Until now, much research has been done on the end-to-end QoS management. Active approaches require the injection of probe packets into the network. The pioneering active approach [29] traceroutes between $37$ participating sites are collected and analyzed to characterize the end-to-end performance issues. The authors of [30] propose to detect path outage among hosts using ping and localizes the observed path outage with traceroute. PlanetSeer [31] relies on active probes to diagnose the root cause of Internet path failures that are detected by passively monitoring the end-users of a content distribution network service deployed on PlanetLab. Commercial systems, e.g., the Keynote [15] and Gomez [16] are also available to detect issues from the end-user¡¯s perspective by active probing. All these work employ active probing while ``Argus'' purely depends on passive monitoring. Although the DiffServ architecture [32] supports end-to-end QoS guarantee, only a limit number of static QoS classes are provided. Obviously, it can not meet the active fine-grained QoS demands of diverse services. Moreover, end-to-end QoS guarantee by DiffServ may require the supports of underlying hardwares, which will introduce overhead to those core devices and result in low efficiency. In addition, fine-grained QoS quantitative guarantee is difficult to be implemented because these fine-grained service anomalies are hard to be traced among ISPs due to commercial security problem.
\subsection{Research Motivation}
In this subsection, we discuss the research motivation from two aspects: (1) the practically observed values of the QoS metrics and contexts, and (2) the causal relationships between them after the training and studying process.
\begin{figure}[htbp]
\centering
\begin{minipage}[t]{0.3\textwidth}
\centering
 \includegraphics[width=1\textwidth,keepaspectratio]{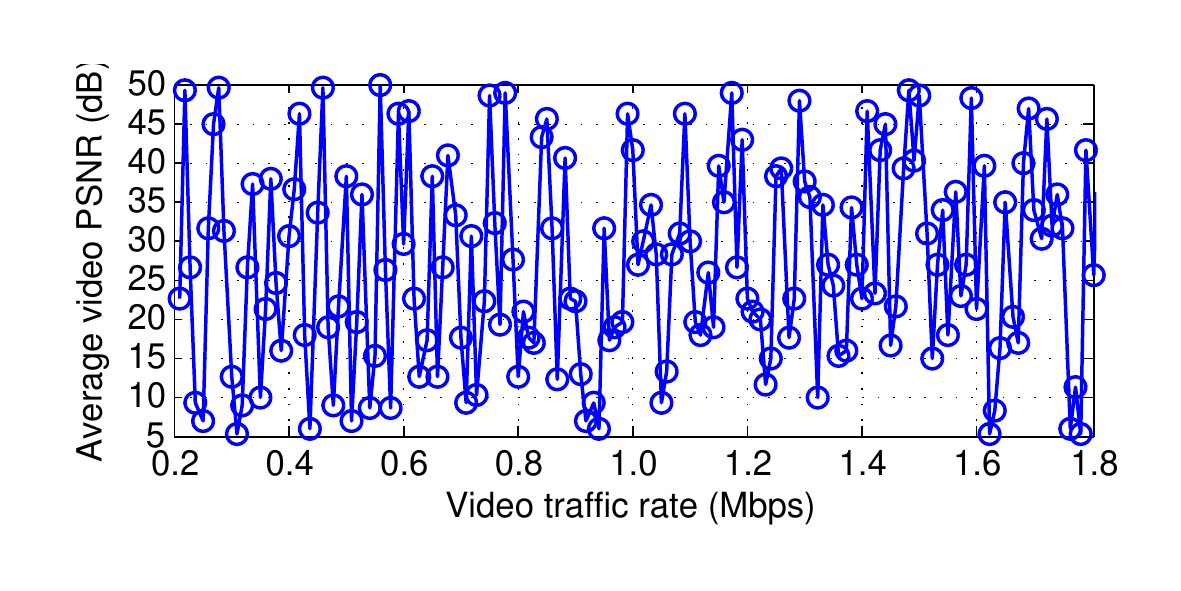}
\end{minipage}
\begin{minipage}[t]{0.3\textwidth}
\centering
 \includegraphics[width=1\textwidth,keepaspectratio]{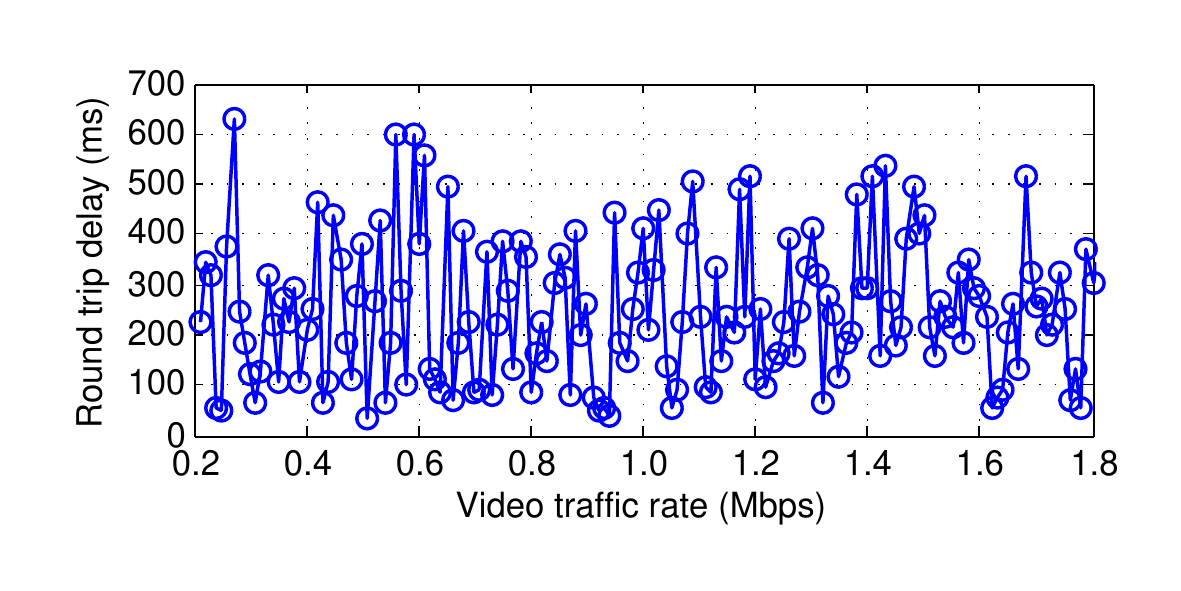}
\end{minipage}
\begin{minipage}[t]{0.3\textwidth}
\centering
 \includegraphics[width=1\textwidth,keepaspectratio]{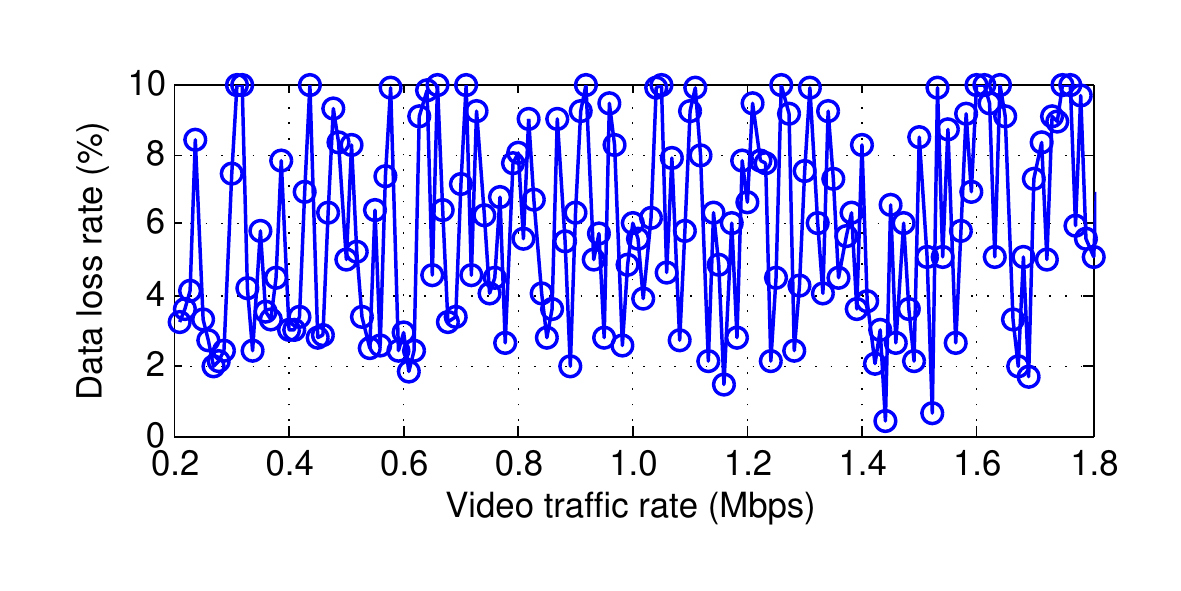}
\end{minipage}
\begin{minipage}[t]{0.3\textwidth}
\centering
 \includegraphics[width=1\textwidth,keepaspectratio]{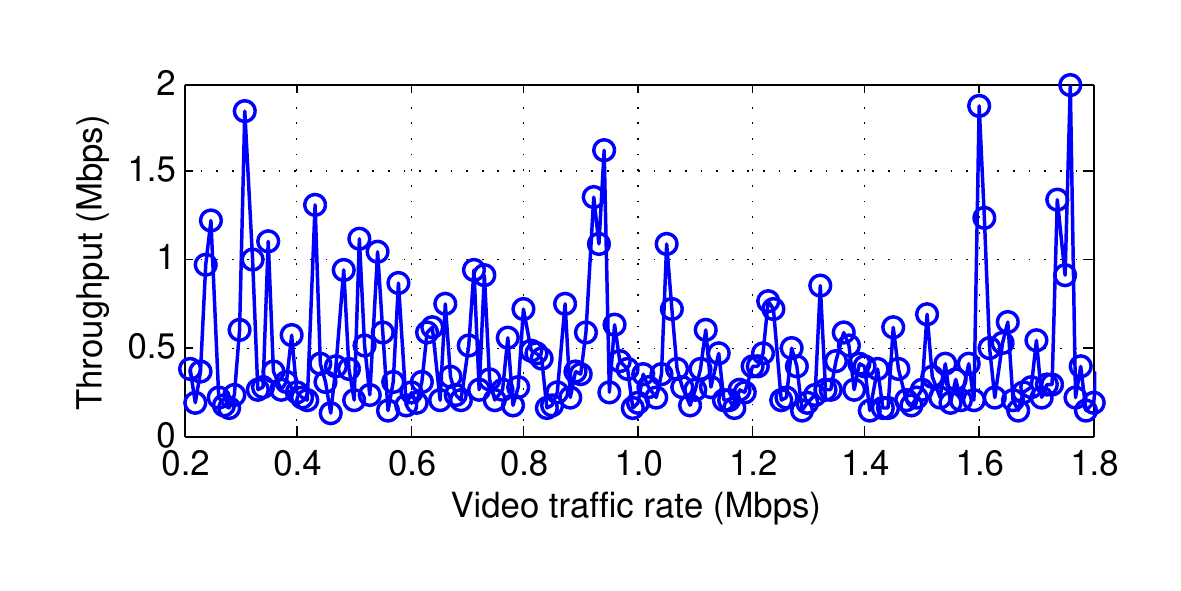}
\end{minipage}
\caption{Profile of the data measurements from our real multimedia conferencing system.}
\end{figure}
\subsubsection{Practically Observed QoS Metrics and Contexts}
First, we depict the real-data measurements of the QoS metrics (video PSNR) and available contexts (video traffic rate, available bandwidth, etc.) from the multimedia conferencing system in Fig. 2. The experimentations lasted for two months and we collected the data from more than $800$ conference sessions for the campus and the Internet users. It can be observed from Fig. 2 that the causal relationships between the video traffic rate and other variables are ambiguous, as no variable changes regularly along with the video traffic rate. To further study the relationships, the raw data needs to be refined and processed. In this work, the Guassian fitting is used to achieve the goal. Next, we shall present and analyze the fitting results.

\subsubsection{Causal Relationships}
After studying on the continuous context samples collected from the experimentations, we found that each context follows an approximate Gaussian distribution, which is a summation of
Gaussian distributions. Due to the lack of space, we only present the video PSNR results in Fig. 3(a) to show that the contexts and QoS metrics in the experiments of this study approximately followed some Gaussian distribution. The other was that some context/QoS metric was approximated or fitted by a sum of Gaussian distributions, instead of one Gaussian distribution. This manner can reduce the fitting residuals as much as possible. In this study, the best fitting was marked out if it had the minimum root mean square error, which is often consistent with the sum of square error.
\begin{figure}[htbp]
\centering
\begin{minipage}[t]{0.35\textwidth}
\centering
 \includegraphics[width=1\textwidth,keepaspectratio]{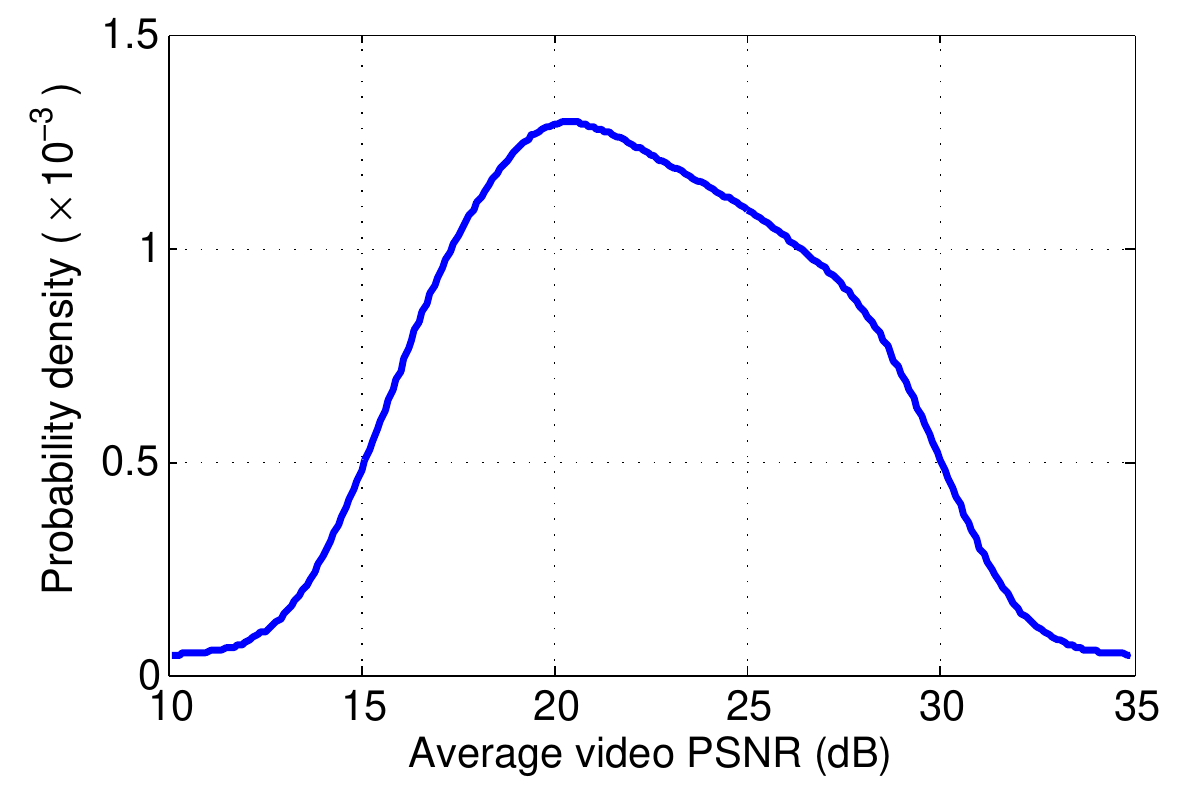}\\
 (a)
\end{minipage}
\quad\quad\quad\quad
\begin{minipage}[t]{0.35\textwidth}
\centering
 \includegraphics[width=1\textwidth,keepaspectratio]{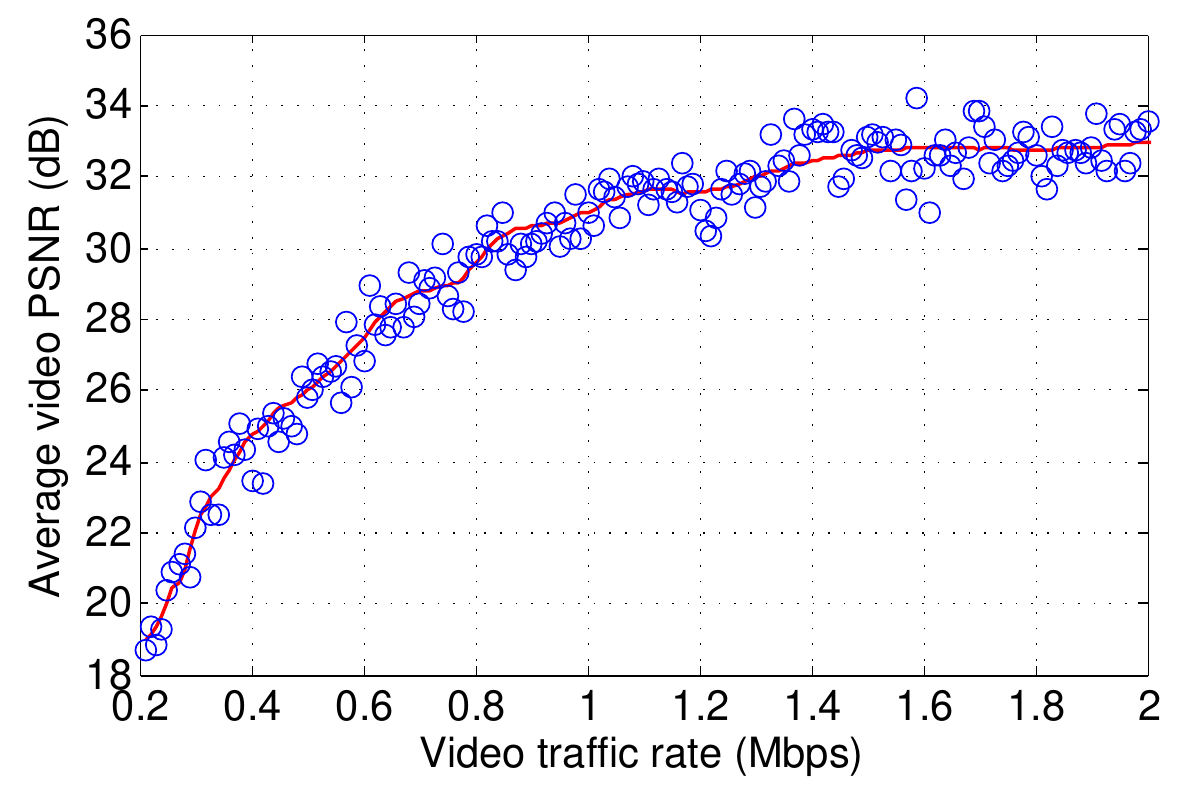}\\
 (b)
 \end{minipage}
\caption{The probability density and relationship: (a) the probability density of average video PSNR after Gaussian fitting, and (b) the causal relationship between them. }
\end{figure}
It should be noted that the probability density distributions in Fig. 3(a) are inconsistent with what are expected to be obtained, since the contexts/QoS metrics are re-arranged according to the increment of video traffic rate depicted in Fig. 2. Fig. 3(b) presents the mapping of the values after the Gaussian fitting process and the relationship between them can be observed. However, these causal relationships represent time-varying characteristics as the video sessions advance. In the next section, we will describe how to determine and exploit the relationships in the proposed solution.
\section{Proposed Solution}
In this section, we outline the over design of the proposed solution and describe the functionality of its major components in detail. The system framework includes working components in both the source and the client side. The information monitor in the client side is responsible for providing the feedback information of all QoS metrics and contexts. The QoS metrics and contexts (e.g., available bandwidth and playback buffer size) can be obtained through specific communication protocols between server and clients. Estimating network status based on end-to-end measurements has been an active research for many years and numerous algorithms have been proposed to achieve the goal. In this paper, the pathChirp [18] algorithm is employed to estimate the end-to-end available bandwidth with high accuracy and efficiency. Next, we will describe the three key phases of CABIN in detail.

\subsection{Information Discretization}
In order to exploit the causal relationships, the feedback QoS metrics/contexts must be properly trained. Fuzzy logic is a multi-valued logic that maps imprecise terms into crisp values. A fuzzy controller is composed of the inference system that includes a rule set, the input membership functions and the output variable. The input values go through a process of fuzzification, where they are converted in terms of the membership functions of the fuzzy sets. These sets are defined over the range of the fuzzy input values, and linguistically describe the variable¡¯s universe of variation. However, fuzzy set theory requires experts to subjectively determine fuzzy sets and their membership functions for a continuous concept. In other words, these fuzzy sets and membership functions may vary from expert to expert. Hence, discretization results may be inconsistent from expert to expert from time to time. To eliminate subjectivity in determining fuzzy sets and membership functions and to assure consistency of discretization result, we propose an adaptive method to discretize a continuous context in this study. Formally, the probability density $f(x)$ of an approximate Gaussian distribution can be defined as:
\begin{equation}
f(x)=\sum_{i=1}^{n}a_i\cdot \exp\left(-\left(\frac{x-b_i}{c_i}\right)^2\right),
\end{equation}
in which $n$ is the total number of sample values, $a_i$ is the coefficient of the $i$th term of the probability density; $b_i$ and $c_i$ are expectation value and standard deviation of the $i$th term, respectively. In this study, a continuous context $c$ is thought to have $n$ discrete values if the probability density of its approximate Gaussian distribution has $n$ terms because each term shows an identical numerical characteristic, i.e., $a_i$, $b_i$ and $c_i$ in the above formal definition. That is, the values covered by a term are more close in semantic than those covered by other terms. Once training data, a set of samples of a target context is collected. The QoS metrics and contexts training algorithm (see Algorithm 1) is referred to make them discrete.
\begin{algorithm}[htbp]
\small
\caption {Information Discretization.}
  \textbf{Input:} Sample values $\mathcal{S}$ of QoS metrics/contexts.
 \\ \textbf{Output:} Discrete value set $DS$ of sample $\mathcal{S}$.
 \\ \textbf{Initialize:}  $\mathbb{D}\Leftarrow {\O},\,\mathbb{M}\Leftarrow {\O},DS\Leftarrow \mathcal{S}$;
 \\ 1: \quad  $\{\mathbb{D},\mathbb{M}\}\Leftarrow$ Dis-Mem$(\mathcal{S})$; /*call the discretization and membership function*/
 \\ 2:  \quad  \textbf{for all} sample $s\in \mathcal{S}$  \textbf{do}
 \\ 3: \quad \quad \quad\quad $f_{D}(s)\Leftarrow \max\{f_{D}(s)|D\in \mathbb{D}, f_{D}\in \mathbb{M}\}$;
 \\ 4: \quad \quad \quad \quad $s\Leftrightarrow D$; /*$s$ is discretized as discrete value $D$*/
 \\ 5: \quad \quad \quad \quad $s\Leftarrow D$;
 \\ 6: \quad  \textbf{end for }
 \\ 7: \quad \textbf{return } $DS$;
 \\ Procedure Dis-Mem($\mathcal{S}$):
 \\ 1: \quad\quad  $PD\Leftarrow$ K-DENSITY($S$); /*estimate the probability density by kernel density estimation*/
 \\ 2: \quad \quad  $\{GF,MSRM\}\Leftarrow$ GAUSSIAN-FITTING($PD$); /*do Guassian fittings on PD*/
 \\ 3: \quad \quad  $\mathbb{G}\Leftarrow GF_{\text{argmin}\{RMSE_i|RMSE_i\in RMSE,1\leq i\leq |RMSE|\}}$;
 \\ 4: \quad \quad  $a_{\max}\Leftarrow 0$;
 \\ 5: \quad \quad  $temp\Leftarrow 0$;
 \\ 6: \quad \quad  \textbf{for each} $t\in \mathbb{G}$ \textbf{do}
 \\ 7: \quad \quad \quad \quad $temp\Leftarrow temp+1$;
 \\ 8: \quad \quad \quad \quad $\mathbb{M}[temp]\Leftarrow t$;
 \\ 9: \quad \quad \quad \quad $\mathbb{D}[temp]\Leftarrow cnt$;
 \\ 10: \quad \, \quad \quad $a_{\max}\Leftarrow \max\{a_{temp},a_{\max}\}$;
 \\ 11: \quad \,  \textbf{end for }
 \\ 12: \quad \, \textbf{for each} $m\in \mathbb{M}$ \textbf{do}
 \\ 13: \quad\quad \quad\quad \quad\,\, $m\Leftarrow \frac{m}{a_{\max}}$; /* normalize the coefficient $a$ (see Equ. (1)) of $m$ by dividing $a_{\max}$*/
 \\ 14: \quad\, \textbf{return } $\{\mathbb{D,M}\}$
 \end{algorithm}
\subsection{Causal Relationship Profiling}
After the available contexts have been trained to be in the right form, the QoS-to-context mapping is performed to find out causal relationships between a QoS metric and these contexts in a real-time manner. A QoS metric is thought to be caused by a context if it changes along with the context, i.e., a causal relationship exists in the QoS metric and corresponding context. The Bayesian network [19] is employed in this paper for studying and modeling these causal relationships.

In the proposed solution, the QoS metric to be studied is considered as a child node, and its contexts are regarded as ancestor nodes. Bayesian network structure learning algorithms are employed here to construct a directed acyclic graph, which qualitatively models the causal relationships between the QoS metrics and the corresponding contexts. Formally, a Bayesian network is a directed acyclic graph with a conditional probability table associated with each discrete node. A conditional probability distribution is associated with a continuous variable. Node $v_a$ is a parent of node $v_b$, which in turn is a child of node $v_a$, if a directed arc from $v_a$ to $v_b$ exists. $G=(V,E)$ denotes the network topology: $V$ denotes the set of nodes and $E$ represents the set of directed arcs. By this catenation connection, the conditional probability table space will be saved exponentially since only parents instead of all ancestors are considered to compute conditional probability. The conditional probability of $v_i$ given its ancestors has been encoded in the probability tables of its parents $p_{v_i}$. For instance, we assume the mean number of discrete values of a node $v_i$ is $e_r$, the mean number of parents is $e_u$ and the mean number of ancestors is $e_a$. Thus, Bayesian network only requires $v_r$ slots instead of $e_r$ to store the conditional probability table for node $v_i$, i.e., the space is saved about $(1-\bar{r}^{u-a})$, which approaches to one exponentially. In order to study the contexts for a target QoS metric using Bayesian network, the QoS metric to be evaluated is considered as a child node, and its corresponding contexts are considered as ancestor nodes. BN structure learning algorithms are referred to construct a direct acyclic graph, which represents the causal relationships between the QoS metric and corresponding contexts. The $K2$ algorithm [20] is applied to learn network structure in this paper. Parents of a node $v$ can only be chosen from the nodes before $v$ in the order. The following three paragraphs convey a node-ordering method.

\begin{figure}[htbp]
\centering
 \includegraphics[width=0.35\textwidth,keepaspectratio]{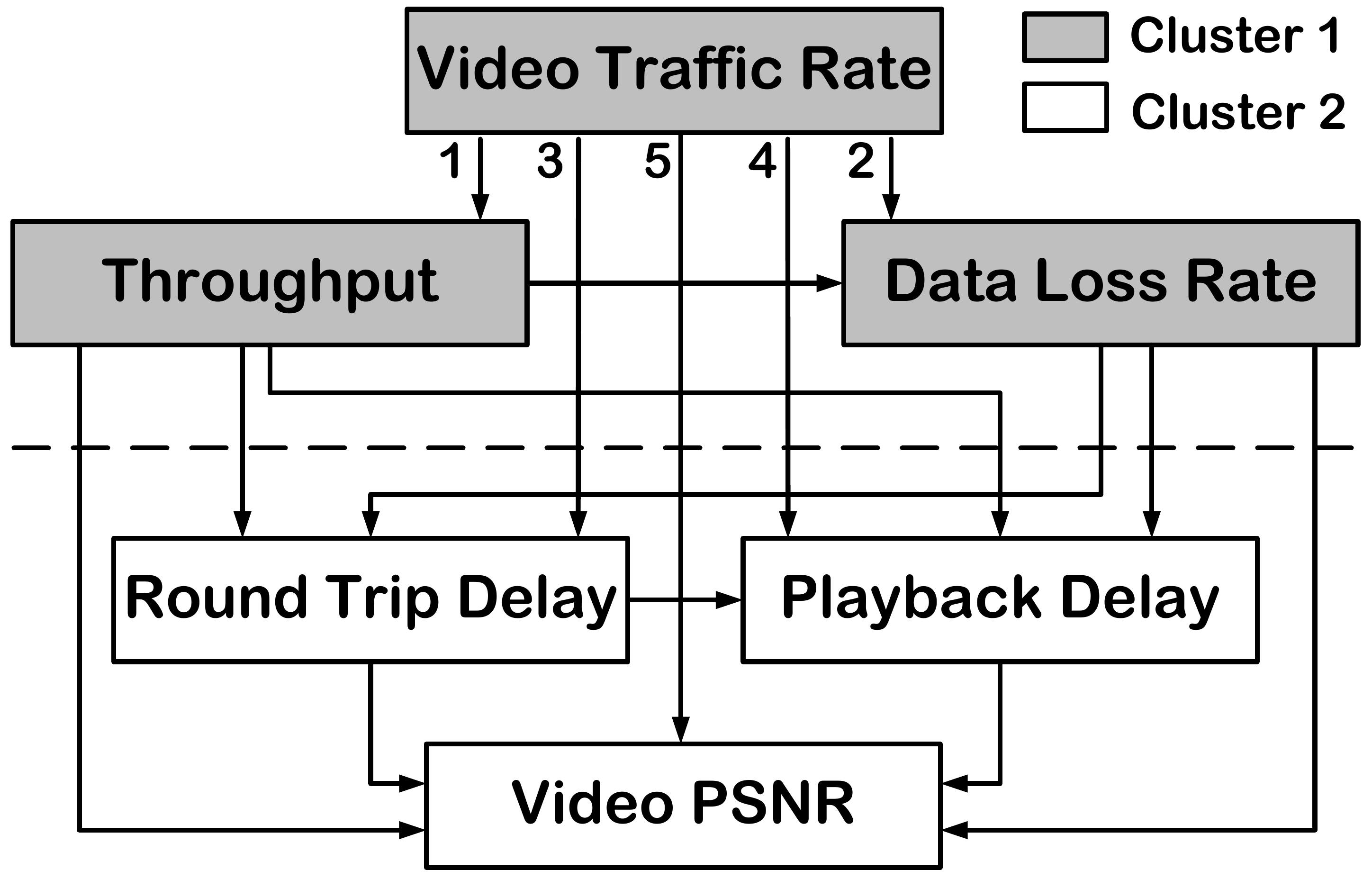}
 \caption{The resulting Bayesian network structure for the QoS-to-context mapping.}\label{9}
\end{figure}
The following steps are taken to exploit the causal relationships between a QoS metric and its contexts. The parent contexts are the causes of a QoS metric. More specifically, parent nodes are direct causes of a child node while other ancestor nodes are the indirect causes. Therefore, the QoS-to-context mapping phase analyzes the relationships in a systematic manner. The resulting Bayesian network structure for the considered QoS metrics/contexts in this paper is presented in Fig. 4. 
\begin{algorithm}[htbp]
\small
\caption {Causal Relationship Profiling.}
\textbf{Step 1.} Get the node order for the $K2$ algorithm with node ordering method described in Section 3.2.\\
\textbf{Step 2.} Form the Bayesian network structure for the QoS metrics and contexts using the $K2$ algorithm.\\
\textbf{Step 3.} For the QoS-to-context mapping, find the parent nodes in the Bayesian network for the QoS metric node by the directed acyclic graph traversal algorithm.
\end{algorithm}
 %
\subsection{Optimal Context Tuning}
Bayesian network parameter training is used to learn the conditional probability tables from training data after the Bayesian network structure is constructed by the the QoS-to-context mapping. In this study, we use Bayesian parameter updating from complete data, i.e., the samples of contexts and QoS metrics. A Bayesian network is completely defined once both its structure and parameters are determined. Then, given the observed data, the Bayesian network inference can be applied to compute the marginal on the specified query nodes. The marginal on a query node $H$ given observed data $E$ can be viewed as the conditional probability $P(H|E)$, which is a quantitative metric. Therefore, we use the marginal on a QoS metric node given the observed contexts to guarantee the QoS metric.

The optimal context adaption is to carry out with a context value that guarantees the QoS metric to the target value with probability $p$. When a user picks up a specific value of the QoS metric, the causal contexts of the QoS metric are tuned to appropriate values with regard to the specific QoS metric value. The optimal context adaption steps are summarized in Algorithm 3.
\begin{algorithm}
\small
\caption {Optimal Context Tuning.}
  \textbf{Step 1.} Find the parent contexts $\mathcal{C}=\{C_{1},C_{2},\ldots,C_{\mathbb{N}}\},1\leq k\leq \mathbb{N},$ for the QoS metric, in which $\mathbb{N}$ is the total number of contexts. The graph is the one learned in the QoS-to-context mapping phase.\\
  \textbf{Step 2.} Exclude the untunable contexts $\mathcal{C}_{P}$ from $\mathcal{C}$ and create a set of tunable parent contexts $\mathcal{C}'=\mathcal{C}-\mathcal{C}_{p}$.\\
  \textbf{Step 3.} Compute the marginal on the query node, the QoS metric, given observed data, the tunable causal contexts.\\
  \textbf{Step 4.} For each discrete value $qm$ of $\mathcal{Q}$, pick conditional probability $P(\mathcal{Q} = q|C_1 = c_1, C_2 = c_2,\ldots, C_m = c_m), C_i \in \mathcal{C}, 1<i< m < N$, with largest value $p$.\\
  \textbf{Step 5.} Adapt the contexts ${C_1,C_2,\ldots,C_m}$ to their corresponding values ${c_1,c_2,\ldots, c_m}$ with regard to $QM = qm$ in step 3. Then, the QoS metric $QM$ can be guaranteed to be the value of $qm$ with probability $p$.
\end{algorithm}
Only parent contexts will be tuned and other contexts will be left untouched. It makes sense by the theory that every set of nodes in a Bayesian network is conditionally independent of QM when conditioned on the Markov blanket QM of the node QM, which consists of its parents, children and the other parents.


\section{Implementation and Evaluation}
In this section, we evaluate the efficacy of CABIN in our real multimedia conferencing system by comparing it with existing QoS guarantee solutions. First, we describe the evaluation methodology that includes the experimental setup, performance metrics, compared approaches and evaluation scenario. Then, we present and discuss the evaluation results in detail.
\subsection{Implementation Issues}
\begin{figure*}[htbp]
\centering
 \includegraphics[width=0.9\textwidth,keepaspectratio]{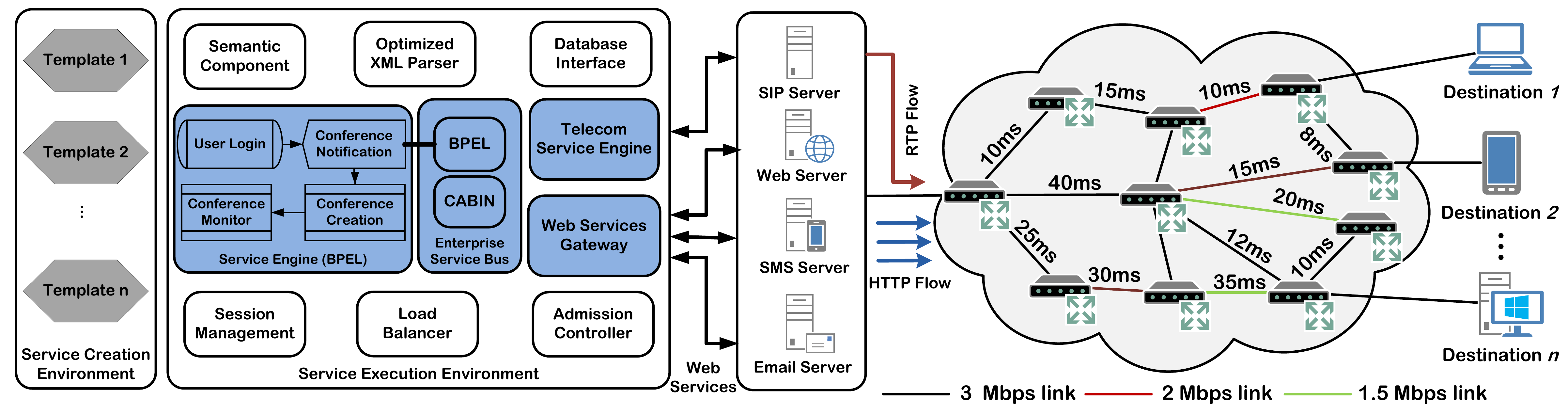}
 \caption{Prototype of multimedia conferencing web services.}\label{9}
\end{figure*}
In this section, we describe the architecture of our multimedia conferencing system and the system implementation of CABIN. Our multimedia conferencing system includes some real-time communication services, e.g., the multiple party call control service and short message service. The system also contains some non real-time Web services, e.g., the user authentication and charging services. Therefore, the system must provide the capability to execute those hybrid services. We have developed the SOA based multimedia conferencing system and the system framework is presented in Fig. 5. Here, the Enterprise Service Bus (ESB) is based on Servicemix [21], which is an open source project. The Telecom service engine kernel is based on another open source project Mobicents [22], which is based on the JAIN SLEE specification and can be easily connected to different application platforms or network elements via a resource adaptor mechanism. In the system design, the BPEL is used for automatically translating the multimedia conferencing processes and properties into a corresponding formal model.

The IBM X3800 8866 serves as the SIP server while the AINOL HD media server and the Radisys CMS9000 are alternate in user for the media process. The conferencing system includes the audio/video services, shared contents and instant messages as depicted in the user interface. Interested readers could refer to [12] for the detailed introductions to our system. Our implementation includes modifications at the media server and the client. The main components of the proposed CABIN are implemented on the Enterprise Service Bus. We significantly extend and modify the system (about $4000$ lines of JAVA code and $2000$ lines of C++ code) to implement the CABIN's components. The function of video traffic rate adaption is realized by the JAVA interface provided by the media server manufacturer.
\subsection{Evaluation Methodology}

\subsubsection{Experimental Setup}
We adopt the Exata [23] as the network emulator. Exata is advanced edition of QualNet [24] in which we can perform semi-physical emulations. In order to implement the realistic emulations of practical network environments, we set the number of nodes and their locations based on the network topology as depicted in Fig. 5. The real computers are connected to the emulation server through the Exata 2.1 connection manager. The parameters of the injected background traffic are listed in Table 1. The video traffic rate is adapted every $2.5$ seconds in all the experiments.
\begin{table}[htbp]
\scriptsize
\setlength{\tabcolsep}{1pt}
\caption{Parameters of background traffic}
\centering
 \begin{tabular}{|c|c|c|c|}
    \hline
   Parameter & FTP & CBR & Pareto \\
    \hline
    \hline
   Start Time (sec) & Rand$[0,100]$ & Rand$[0,50]$ & Rand$[50,150]$\\
    \hline
    Duration (sec) & Rand$[0,300]$ & Rand$[50,200]$ & Rand$[200,300]$\\
    \hline
    Packet size (Bytes) & $1500$ & $500$ & $1000$\\
    \hline
    Traffic Rate (Mbps) & N/A & Rand$[1-1.5]$ & Rand$[0.5-1]$\\
    \hline
    Data Size & Rand$[10,1500]$KB & Rand$[5,50]$MB & N/A \\
    \hline
  \end{tabular}
\end{table}

\subsubsection{Performance Metrics}
\begin{itemize}
  \item PSNR (Peak Signal-to-Noise Ratio) is a standard metric of video quality and is a function of the mean square error between the original and the received video frames. If a video frame is dropped or past the deadline, it is considered to be lost and is concealed by copying from the last received frame before it.
  \item Playback delay. The playback delay process corresponds to the time left to playback the video data at the tail of the playback buffer or, equivalently, the buffered video length plus the time left to start video playback.
  \item Throughput. The throughput is the bandwidth used for the video application and it represents the capability of the competing approaches to effectively exploit the available bandwidth. This network-level metric is measured by the Exata Analyzer.
\end{itemize}
\subsubsection{Compared Approaches}
We compare the proposed solution with the following context-aware approaches:
\begin{itemize}
  \item Throughput oriented approach [25-26]. The throughput oriented (TON) adaption approaches aim at maximizing the application's throughput by adapting the video traffic rate. In this work, we assume the TON scheme adjusts the video encoding rate according to the average `loss-free' bandwidth of all the participants.
  \item Delay oriented approach [27-28]. The delay oriented (DON) scheme adjusts the video traffic rate based on the feedback info of playback buffer size from the clients. Playback buffer starvation occurs when a video frame is requested by the application layer for playback before being fully received. From a user QoS perspective, playback buffer starvation is undesirable because it results in video frame freezes and re-buffering. Therefore, the constraint for the video traffic rate is equal to prevent the buffer underflow.
\end{itemize}
\subsubsection{Evaluation Scenario}
In order to fully evaluate the performance of the compared approaches, we conduct experiments with different numbers of participants: $4$, $8$, $12$ and $16$. For the standard deviation results, we repeated each round of experimentations more than $5$ times and obtain the average values with a $95\%$ confidence interval.
\subsection{Experimental Results}
\subsubsection{PSNR}
As depicted in Fig. 6, CABIN achieves higher average PSNR values and lower deviations than the TON and DON schemes in all the evaluation scenarios. It can be observed that the superiority of CABIN over the competing schemes becomes more distinct as the number of participants increases. The results indicate the relationships between the QoS metrics and corresponding contexts become more complex in case of many online users. The average video quality degrades with the increase in the number of users for all the approaches. Indeed, due to the capability limitations of the media server, the user-perceived video quality is a challenging problem to be tackled in our future work. In general, CABIN improves the average video PSNR by up to $3.03$ and $4.16$ dB compared to the TON and DON scheme and the video PSNR per frame is shown in Fig. 7.
\begin{figure}[htbp]
\centering
\begin{minipage}{1\linewidth}
\centering
 \includegraphics[width=0.7\textwidth,keepaspectratio]{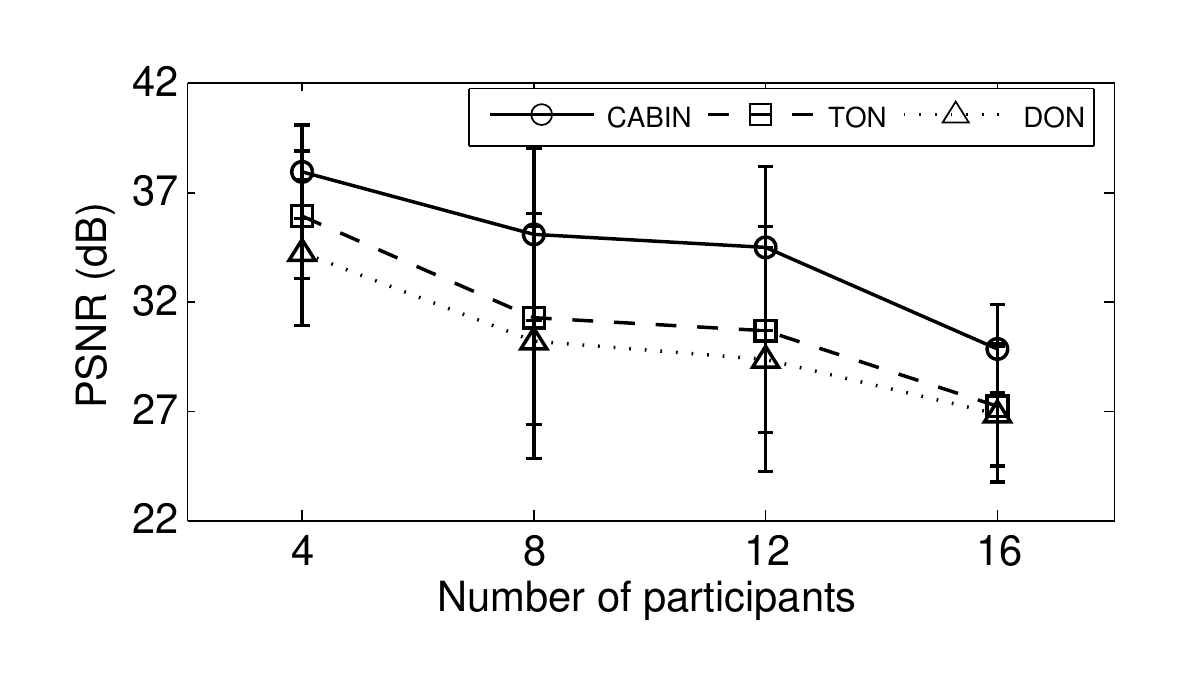}
 \centering
 \caption{Average PSNR values for all the competing approaches in different experimental scenarios.}\label{9}
\end{minipage}%

\begin{minipage}{1\linewidth}
\centering
 \includegraphics[width=0.7\textwidth,keepaspectratio]{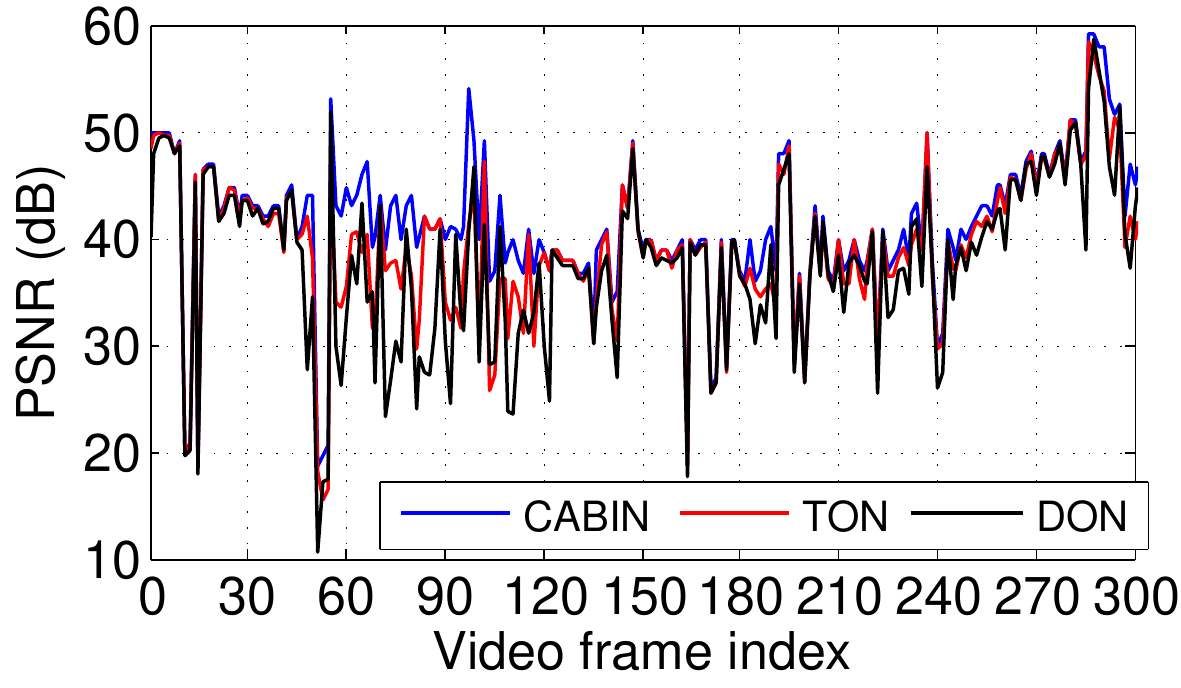}
 \centering
 \caption{Instantaneous PSNR value of all the competing approaches for each video frame indexed from $1$ to $300$.}\label{9}
\end{minipage}%
\end{figure}
\subsubsection{Playback delay}
The playback delay is defined as the time left to playback the video data at the tail of the playback buffer. In Fig. 8, we plot the average playback delay and DON maintains the lowest latency of all the competing approaches. In order to have a microscopic view of the results, we show the playback delay process of $[0,250]$ seconds in Fig. 9. It can be observed from Fig. 9 that the DON is able to regulate the playback buffer contents based on its size by adjusting the video traffic rate to maintain enough data to compensate for the time-varying networks status. Furthermore, it avoids the playback buffer starvation and re-buffering. The buffering behavior is also correlated with the available bandwidth. Although the performance of CABIN is generally inferior to that of DON, it still exhibits good performance in preventing the playback buffer starvation.
\begin{figure}[htbp]
\centering
\begin{minipage}{1\linewidth}
 \includegraphics[width=0.7\textwidth,keepaspectratio]{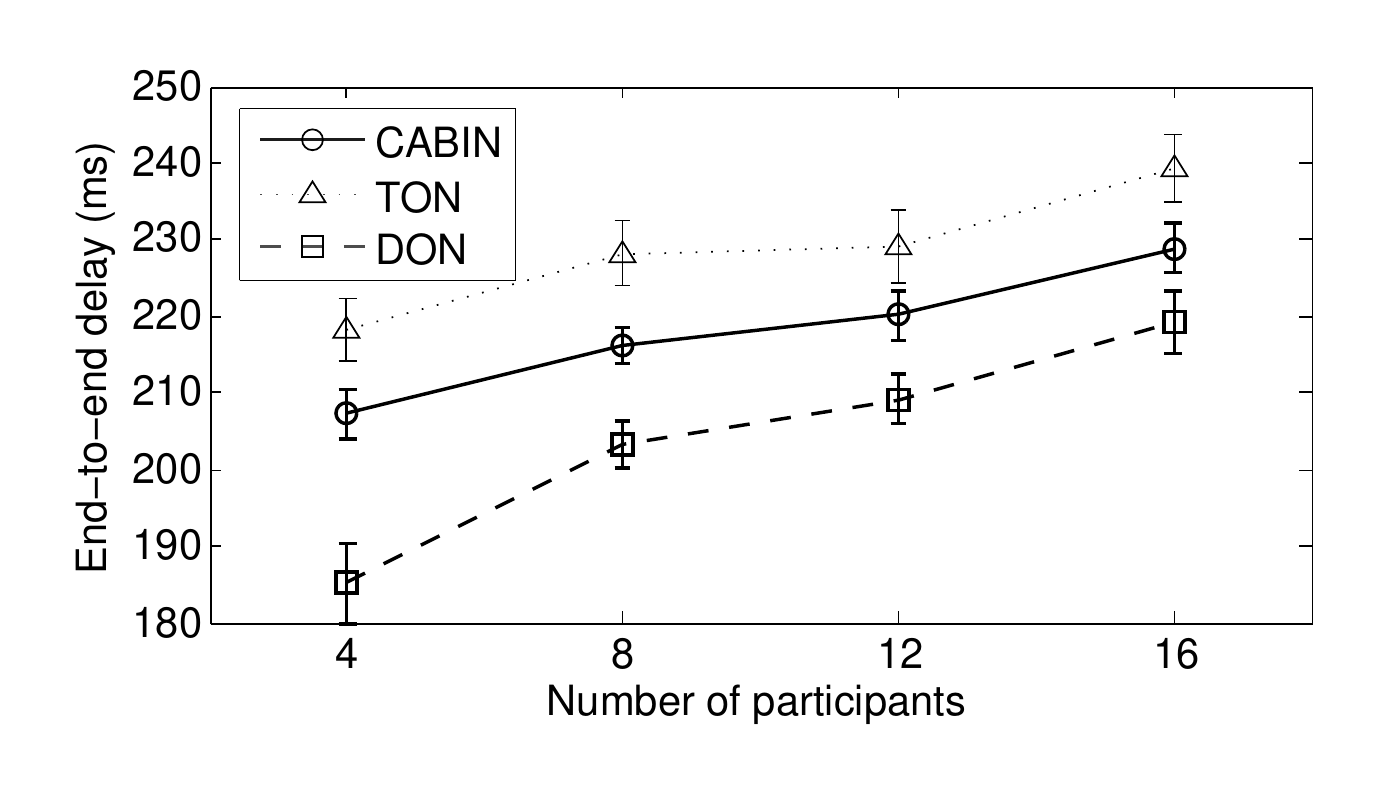}
 \centering
 \label{9}
\end{minipage}%
\caption{Average playback delay of all the participants in the experimental scenarios.}
\begin{minipage}{1\linewidth}
\centering
 \includegraphics[width=0.7\textwidth,keepaspectratio]{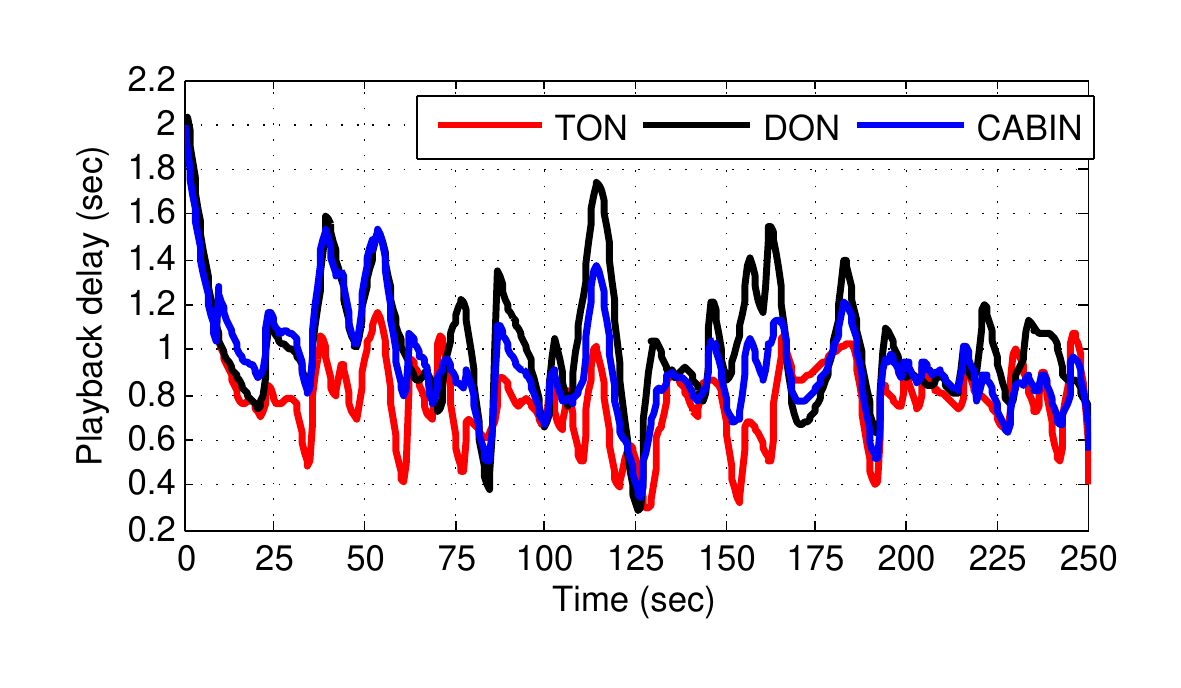}
 \centering
 \caption{Instantaneous value of playback delays during the interval of $[0, 250]$ sec.}\label{9}
\end{minipage}%
\end{figure}
\subsubsection{Throughput}
Fig. 10 plots the average throughput values for all the approaches under different scenarioes. The TON approach achieves the highest throughput as it generally adjusts the video traffic rate to larger values than the competing schemes. However, the real-time video applications can not directly benefit from the throughput gains as the end-to-end delay increases simultaneously. For real-time video applications, delay is another key performance since each video packet is associated with a decoding deadline. The overdue packets will be discarded at the client side since they can no longer contribute to the decoding process. Fig. 11 illustrates the instantaneous throughput values of all the competing schemes during the interval of $[0,300]$ second.
\begin{figure}[htbp]
\centering
\begin{minipage}{1\linewidth}
 \includegraphics[width=0.7\textwidth,keepaspectratio]{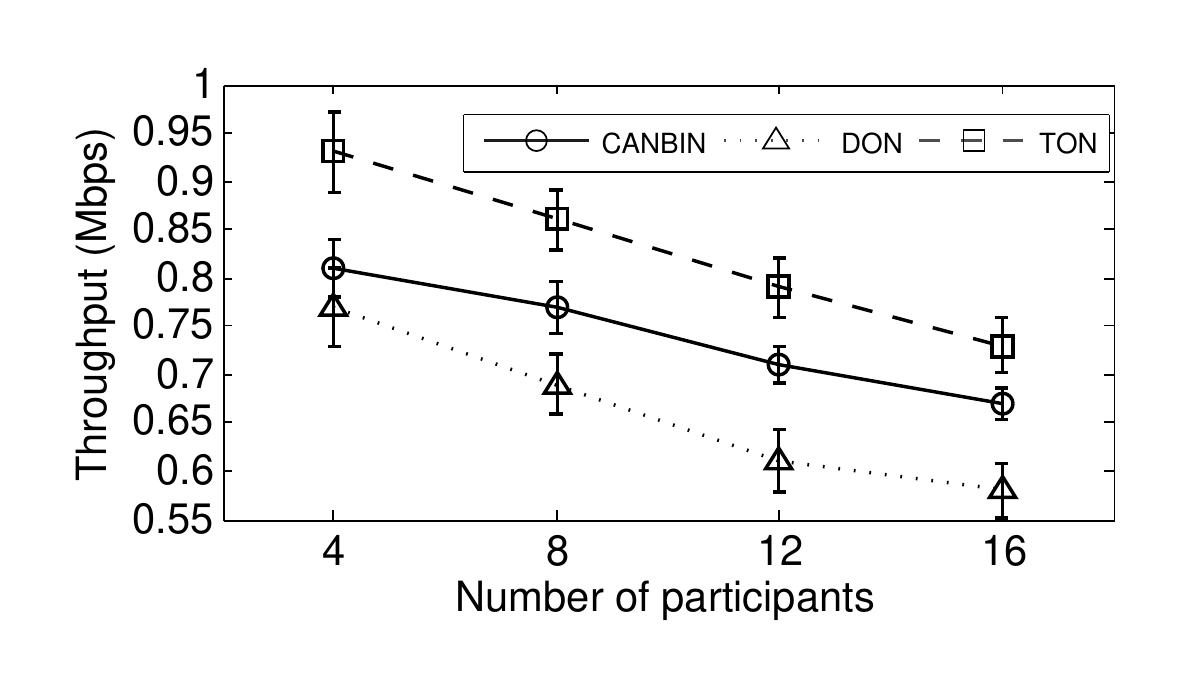}
 \centering
\label{9}
\end{minipage}%
\caption{Average throughput values for all the competing approaches in different evaluation scenarios.}
\begin{minipage}{1\linewidth}
\centering
 \includegraphics[width=0.7\textwidth,keepaspectratio]{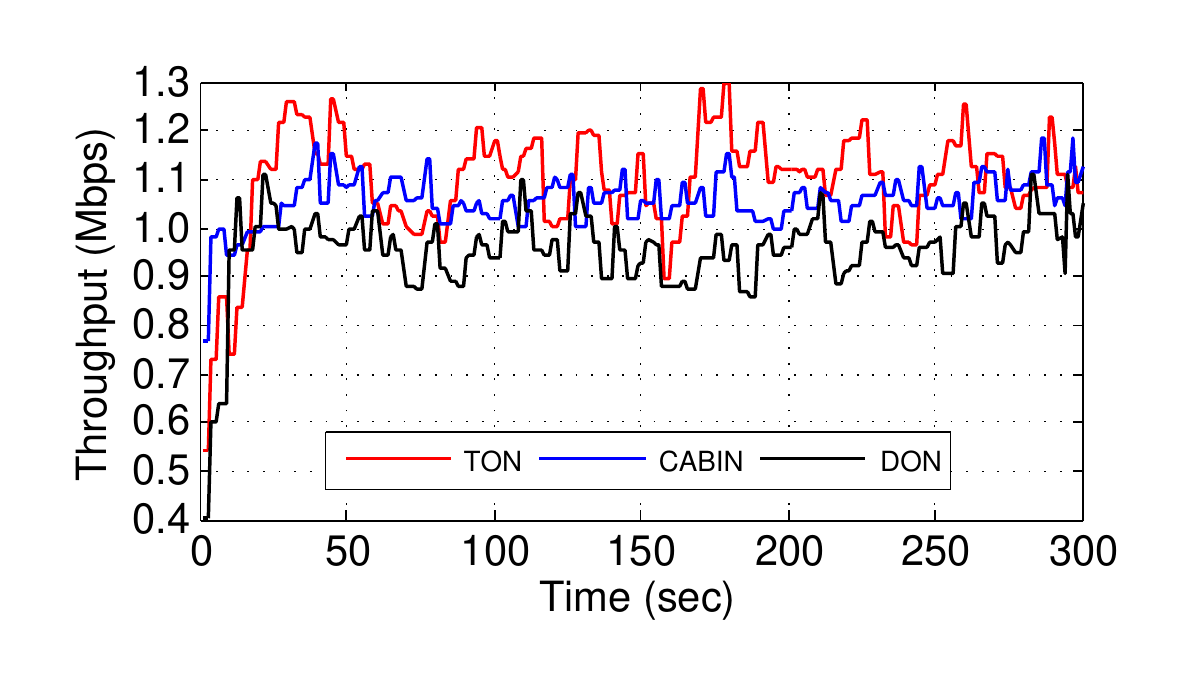}
 \centering
 \label{9}
\end{minipage}%
\caption{The instantaneous throughput values for all the competing approaches during the interval of [0,300] second.}
\end{figure}
\subsection{Discussion}
The evaluation results demonstrate that the proposed CABIN outperforms the TON and DON schemes in improving the target QoS metric (i.e., the video PSNR). Due to the lack of space, we do not present the results of computational cost of CABIN and its competing models in this section. Indeed, the performance gain of CABIN is at the sacrifice of computation and network overhead. However, with the advances in the system hardware and software, this overhead is negligible comparing to the service creation and execution time. On the other hand, the major flaw with both TON and DON schemes is that they are highly dependent on one specific context. Therefore, the accuracy of obtained contexts imposes significant impact on their performances. It is well known the networks status monitoring is still a challenging problem remains to be largely explored. As CABIN takes all the available contexts into consideration in a comprehensive manner, it is less single-context-dependent and thus achieves better results.
\section{Conclusion}
In this paper, we have presented the CABIN, a hybrid web and communication service framework to comprehensively exploit the causal relationships between the QoS metrics and corresponding contexts. CABIN offers a one-stop-shop integrating web and communication services, and a flexible scheme to orchestration them. The performance of CABIN is evaluated in our real multimedia conferencing system [35][36] and experimental results demonstrate its superiority over existing approaches. As future work, we will consider evaluating the performance of CABIN in other multimedia conferencing services [45], e.g., the audio service.





%

\end{document}